\documentclass{article}

\usepackage{amssymb,amsfonts,amsmath,stmaryrd}
\usepackage{cite,enumerate,float,indentfirst}
\usepackage{color}
\usepackage{tikz}
\usetikzlibrary{arrows,snakes,backgrounds}
\usepackage{url}
\usepackage{hyperref}
\usepackage{textcomp}

\def\be{\begin{eqnarray}}
\def\ee{\end{eqnarray}}
\def\nn{\nonumber}

\definecolor{red}{rgb}{1,0,0}
\definecolor{orange}{rgb}{1,0.5,0}
\definecolor{violet}{rgb}{0.7,0,1}

\hoffset= - 1.0in         % switch off for draft style
\begin{document}

\title{\vspace{-1cm}{\Large {\bf
      Harer-Zagier formulas for knot matrix models
    }
    \date{}
    \author{
      {\bf A. Morozov$^{a,b,c,*}$}
      {\bf A. Popolitov$^{a,b,c,\dag}$}
      {\bf Sh. Shakirov$^{b,\ddag}$}}
}}

\maketitle
\vspace{-4.2cm}

\begin{center}
	\hfill ITEP/TH-03/21 \\
	\hfill IITP/TH-03/21 \\
	\hfill MIPT/TH-03/21
\end{center}

\vspace{1.7cm}

\begin{center}
  $^a$ {\small {\it Institute for Theoretical and Experimental Physics, Moscow 117218, Russia}}\\
  $^b$ {\small {\it Institute for Information Transmission Problems, Moscow 127994, Russia}}\\
  $^c$ {\small {\it Moscow Institute of Physics and Technology, Dolgoprudny 141701, Russia }} \\
  \vspace{0.25cm}
  $*$ {\small {\it morozov.itep@mail.ru}} $\dag$ {\small {\it popolit@gmail.com}} $\ddag$ {\small {\it shakirov.work@gmail.com}}
\end{center}

\vspace{0.5cm}

\begin{abstract}
  Knot matrix models are defined so that the averages of
  characters are equal to knot polynomials.
  From this definition one can extract single trace
  averages and generation functions for them in the
  group rank -- which generalize the celebrated
  Harer-Zagier formulas for Hermitian matrix model.
  We describe the outcome of this program for HOMFLY-PT
  polynomials of various knots.
  In particular, we claim that the Harer-Zagier formulas
  for torus knots factorize nicely, but this does not happen
  for other knots.
  This fact is mysteriously parallel to existence of
  explicit $\beta=1$ eigenvalue model construction for torus knots only,
  and can be responsible for problems with construction of a similar
  model for other knots.
\end{abstract}

\bigskip

\section{Introduction}

Superintegrability property usually means that
some {\it complete} set of averages is explicitly calculable.
The famous examples of this kind begin with harmonic oscillator and
the motion in Coulomb potential.
Eigenvalue matrix models \cite{
  % UFN3
  paper:M-matrix-models-as-integrable-systems,
  paper:M-challenges-of-matrix-models,
  paper:M-quantum-deformations-of-tau-functions
} seem to possess this property
in the following sense \cite{
  % IMM
  paper:IMM-ward-identities-and-combinatorics-of-rainbow-tensor-models,
  paper:IMM-tensorial-generalization-of-characters,
  paper:IMM-complete-solution-to-gaussian-tensor-model,
  % MM
  paper:MM-on-the-complete-perturbative-solution-of-on-matrix-models,
  paper:MM-sum-rules-for-characters-from-characters-preservation-property-of-matrix-models
}: averages of characters are explicitly known -- they are again characters
\cite{
  paper:IMMM-character-expansion-for-homfly-polynomials-III,
  % MSh
  paper:MMS-proving-agt-as-hs-duality
}.
Moreover, the dependence on the size of the matrix $N$ is captured in the
{\it topological locus} -- the specialization of time-variables at the r.h.s.
of the relation $<character>\ \sim character$.
Therefore the $N$-dependence is polynomial in $q^N$ for $q$-deformed models
and just polynomial in $N$ in the limit $q\longrightarrow 1$.
Laplace transform in $N$ converts this average into a rational function.

Remarkably, sometimes  there is even more:
not only denominator, but numerator is drastically simplified --
all the roots are just plus-or-minus powers of $q$.
However, this happens for single-trace averages $<P_k>$ rather than characters
and one needs to derive a special formula to describe the answer.
We call these expressions the {\it Harer-Zagier formulas} (HZF), because they were first
discovered in \cite{
  % HZ
  paper:HZ-the-euler-characteristic-of-the-moduli-space-of-curves
} for the simplest Gaussian Hermitian model
(see  \cite{
  % MSh_hz,
  paper:MSh-exact-2-point-function-in-hmm,
  % MSh_bh
  paper:MSh-from-brezin-hikami-to-harer-zagier-formulas-for-gaussian-correlators
} for further developments in that case,
including relation to other interesting subjects like Brezin-Hikami formulas \cite{
  paper:BH-duality-and-replicas-for-a-unitary-matrix-model
},
Okounkov's exponentials \cite{
  % Ok
  paper:O-generating-functions-for-intersection-numbers-on-moduli-spaces-of-curves
} and $W$-representations
\cite{
  % MSh_Wrep
  paper:MSh-generation-of-matrix-models-by-w-operators
}).
Reversing the statement, the  claim is that
\be
\text{HZF for single-trace averages are rational functions with zeroes and poles at plus-or-minus powers of\ } q
\label{statement}
\ee
This is a frequent property of eigenvalue matrix models and it can serve
as an alternative manifestation of superintegrability phenomenon.
An important question is what is the relation between the two manifestations,
and, if different, which is more restrictive.

In this paper we consider the first example when the difference occurs --
the hypothetical {\it knot} matrix models, where superintegrability is assumed to imply
that averages of characters are the corresponding colored HOMFLY-PT polynomials.
What we demonstrate is that this implies factorization of the single-trace averages
in the spirit of (\ref{statement}) \textit{only} for torus knots and only for the $Sl_N$ --
which so far remain the only case, when the eigenvalue formula
(the $\beta=1$ TBEM model \cite{paper:T-soft-matrix-models-and-chern-simons-partition-functions,
  paper:BEM-torus-knots-and-mirror-symmetry
}) is actually known.
Put differently, the statement is that the $\beta=1$
{\it TBEM  matrix model for torus knots does possess the property (\ref{statement})},
while (\ref{statement}) does not follow from assumption
\be
<character> \ =
\ HOMFLY-PT
\label{suknots}
\ee
for non-torus knots.
In fact our result questions the relevance of the postulate \eqref{suknots}
beyond the torus-knot variety, i.e. implies that
\textbf{the problem of knot matrix models remains open.}

%{\footnotesize
%At the moment the proof that (\ref{statement}) holds for torus knots is pure technical
%and is actually equivalent to computer calculation.
%Actually, such was also the proof of the same property for $q$-deformed Gaussian model
%in \cite{
%  % MPSh
%  paper:MPSh-quantization-of-harer-zagier-formulas
%}.
%Some more conceptual argument is still needed in both cases.
%Also of interest is the close similarity between the %Harer-Zagier functions
%in these two cases -- it is largely inplied by (\ref{statement}), still a less
%technical argument is highly desirable.}

\section{Knot matrix models}

Matrix model for a knot ${\cal K}$ can be {\it defined} by the
hypothetical/desirable superinterability property
\cite{
  % MM,
  paper:MM-on-the-complete-perturbative-solution-of-on-matrix-models,
  % MMsr,
  paper:MM-sum-rules-for-characters-from-characters-preservation-property-of-matrix-models,
  % MMlast
  paper:MM-superintegrability-and-kontsevich-hermitian-relation
}
\be
\left<\chi_R\right>^{\cal K}\ := \ {\cal P}_R^{\cal K}(q,A=q^N)
\label{siknots}
\ee
This definition is supported by existence of explicit TBEM  model \cite{
  % T,
  paper:T-soft-matrix-models-and-chern-simons-partition-functions,
  % BEM
  paper:BEM-torus-knots-and-mirror-symmetry
}
for torus knots, where the measure at the l.h.s. contains
peculiar deformations of triginometric Vanderomonde functions
(see sec.\ref{torus} below).
Equally explicit formulas are not yet available even for twisted knots
(see \cite{
  % AMMM
  paper:AMMM-towards-matrix-model-representation-of-homfly-polynomials
} for explanation of related beauties and difficulties),
but eq.\eqref{siknots} allows to bypass them.
In fact, definition/ansatz \eqref{siknots} is very restrictive and allows
for a number of non-trivial consistency checks -- a possibility, which we
begin to explore in the present paper.

On the other hand, ansatz \eqref{siknots} has some freedom.
The characters $\chi_R$ can be chosen as \textit{various} symmetric polynomials
(Schur, Jack, Macdonald etc.) and ${\cal P}_R^{\cal K}$ can be chosen to be
\textit{various} knot polynomials (HOMFLY-PT, Khovanov-Rozansky, super- or hyper-polynomials etc.)
in a suitably chosen \textit{framing}.

All these details are crucial to the transition \eqref{siknots} $\rightarrow$ \eqref{statement},
and so current lack of explicit eigenvalue model representation for knot matrix models beyond
torus knots can be, at least partly, attributed to difficulty
of fixing these freedoms \textit{simultaneously} in exactly the right way.

In cases when characters $\chi_R$ are Schur functions (i.e. for $\beta=1$ eigenvalue models), the single-trace operators (times) are
expressed entirely through single-hook characters:
\be
P_k = \sum_{i=0}^{k-1} (-)^i {\rm Schur}_{[k-i,1^{i}]}\{P\}
\label{projk}
\ee
while the simplest possibility for knot polynomials \cite{
  % knotpols
  paper:A-topological-invariants-of-knots-and-links,
  paper:C-an-enumeration-of-knots-and-links,
  paper:J-index-for-subfactors,
  paper:J-a-polynomial-invariant-for-knots-via-von-neumann,
  paper:J-hecke-algebra-representations-of-braid-groups-and-link-polynomials,
  paper:K-state-models-and-the-jones-polynomial,
  paper:HOMFLY-a-new-polynomial-invariant-of-knots-and-links,
  paper:PT-invariants-of-links-of-conway-type
} are HOMFLY-PT polynomials.
This implies that
\be \label{eq:pk-through-schur}
\left<P_k\right>^{\cal K}_N = \sum_{i=0}^{k-1} (-)^i {\cal H}^{\cal K}_{[k-i,1^{i}]}(q,q^N)
\ee
One can further perform a Laplace transform in $N$ to get a (1-point) Harer-Zagier function
\be
{\cal Z}^{\cal K}_k := \sum_{N=0}^\infty \lambda^N\cdot \left<P_k\right>^{\cal K}_N
\ee
Also various generation functions w.r.t. the $k$-variables can be introduced,
but while they lead to simplification at $q=1$, they seem to blur matters in the $q$-deformed case
\cite{
  % MPSh
  paper:MPSh-quantization-of-harer-zagier-formulas
}.

\section{Torus knots matrix model
\label{torus}}

TBEM  model \cite{
  % T,
  paper:T-soft-matrix-models-and-chern-simons-partition-functions,
  % BEM
  paper:BEM-torus-knots-and-mirror-symmetry
} provides explicit measure in (\ref{siknots})
for torus knot ${\cal K} = {\rm Torus}_{m,n}$ and for the gauge group $SU(N)$
(HOMFLY-PT invariants \cite{
  % knotpolynomials
  paper:A-topological-invariants-of-knots-and-links,
  paper:C-an-enumeration-of-knots-and-links,
  paper:J-index-for-subfactors,
  paper:J-a-polynomial-invariant-for-knots-via-von-neumann,
  paper:J-hecke-algebra-representations-of-braid-groups-and-link-polynomials,
  paper:K-state-models-and-the-jones-polynomial,
  paper:HOMFLY-a-new-polynomial-invariant-of-knots-and-links,
  paper:PT-invariants-of-links-of-conway-type
}):
\be
\left< F \right>^{{\rm Torus}_{m,n}}
\sim
\oint_{\substack{\text{unit} \\ \text{circle}}} F\{e^{x_i}\} \prod_{i<j}^N \sinh \frac{x_i-x_j}{m} \sinh \frac{x_i-x_j}{n}
\,\prod_{i=1}^N e^{-x_i^2/(2g)}\,dx_i
\label{avtorus}
\ee
where $q=e^{\frac{g}{2mn}}$, then (\ref{siknots}) becomes a non-trivial theorem
(proved in above mentioned papers).
Generalization of this explicit formula for other knots is still an
open problem (see \cite{
  % AMMM
  paper:AMMM-towards-matrix-model-representation-of-homfly-polynomials
} for discussion).
For our purposes in this paper we can use just (\ref{siknots}), even if the measure is unknown.
However, for torus knots this consideration is better grounded since (\ref{avtorus})
provides an explicit realization of the average $\left<\ldots\right>$.

In the torus case we can define HOMFLY-PT polynomials as functions of arbitrary time variables
\cite{
  % DMMSS
  paper:DBMMSS-superpolynomials-for-torus-knots-from-evolution-induced-by-cut-and-join-operators
}:
\be
{\cal H}_R^{{\rm Torus}_{m,n}}\{p\}
\sim q^{\frac{2n\hat W_2}{m}}\cdot  {\rm Schur}_R\{p_{mk}\}
= \sum_{Q\in m|R|} q^{\frac{2n\varkappa_Q}{m}}\cdot C_R^Q\cdot  {\rm Schur}_Q\{p_k\}
\label{torfor}
\ee
where $\varkappa_Q:=\sum_{(i,j)\in Q} (j-i)$
and Adams coefficients $C_R^Q$ describe expansion of Schur functions
\be
 {\rm Schur}_R\{p_{mk}\}  = \sum_{Q\in m|R|}   C_R^Q\cdot  {\rm Schur}_Q\{p_k\}
\label{adams}
\ee
These polynomials become topological invariants, in particular, acquire $m-n$ symmetry,
when time variables are restricted to topological locus:
\be
p_k = p_k^* := \frac{A^k - A^{-k}}{q^k-q^{-k}}
\ee
For the particular gauge group $SL(N)$ the variables $A=q^N$, so that $p_k^* = \frac{[kN]}{[k]}$
becomes a ratio of $q$-numbers $[x]:=\frac{q^x-q^{-x}}{q-q^{-1}}$, and we get
\be
\left< {\rm Schur}_R\right>^{{\rm Torus}_{m,n}}_{SL(N)} = A^{n|R|}
\ \sum_{Q\in m|R|} q^{\frac{2n\varkappa_Q}{m}}\cdot C_R^Q \cdot
{\rm Schur}_Q\left\{\frac{[Nk]}{[k]}\right\}
\label{aveSchur}
\ee
where proportionality factor $A^{n|R|}$ ensures agreement with (\ref{avtorus}). We can finally convert them into Harer-Zagier functions
\be \label{eq:laplace-transform}
{\cal Z}^{{\rm Torus}_{m,n}}_{k} :=
\sum_{N=0}^\infty \lambda^N\cdot \left< P_k\right>^{{\rm Torus}_{m,n}}_{SL(N)}
\ee
At this point it is important to comment on the framing of HOMFLY polynomials here. It is often convenient to deal with knot polynomials which are reduced and are in \emph{topological} framing -- then they satisfy simple differential-expansion identities like $\overline{{\cal H}_{[r]}}(q,A) -1 \sim \{Aq^r\}\{A/q\}$  in symmetric representations,
where overline dentes reduced knot polynomial ${\cal H}_R = D_R\cdot \overline{{\cal H}_R}$
and $D_R$ is quantum dimension of representation $R$. However the framing of HOMFLY polynomials obtained here from TBEM model is generally not topological: instead it is so-called \emph{vertical} or \emph{spectral} framing.

\section{Fundamental representation $R=[1]$}

The ansatz \eqref{siknots},\eqref{eq:pk-through-schur} becomes very simple for the first single-trace average
$\left \langle p_1 \right \rangle$. Namely, only the fundamental character $\chi_\Box$ does contribute.
This provides a quick test for various options to insert in ansatz \eqref{siknots}, as the Harer-Zagier factorization
\eqref{statement}, if present, should occur already at this level.
In this section we evaluate different possibilities using this test.

Let us begin with the simplest case non-normalized HOMFLY-PT, where the answer for $\left \langle p_1 \right \rangle$
is given by a single fundamental HOMFLY-PT
\be
{\cal Z}^{{\rm Torus}_{m,n}}_{1} :=
\sum_{N=0}^\infty \lambda^N\cdot {\cal H}^{{\rm Torus}_{m,n}}_{[1]}
\ee
For example, in the case of the trefoil $(m,n)=(2,3)$ the fundamental HOMFLY-PT is
\be
{\cal H}^{{\rm Torus}_{2,3}}_{[1]} = A^6 \frac{\{A\}}{\{q\}}\left(1 - A^{-2}\{Aq\}\{A/q\}\right)
= \frac{(q^2+q^{-2})q^{5N} - (q^2+1+q^{-2})q^{3N} + q^{N}}{q-q^{-1}}
\ee
and we get a Harer-Zagier formula
\be
{\cal Z}^{{\rm Torus}_{2,3}}_{1}
=\frac{1}{q-q^{-1}}\left( \frac{q^2+q^{-2}}{1-q^5\lambda}
  - \frac{q^2+1+q^{-2}}{1-q^3\lambda} + \frac{1}{1-q\lambda}\right)
= \frac{\lambda q^3 (q^3-\lambda)}{(1-q\lambda)(1-q^3\lambda)(1-q^5\lambda)}
\ee
which is a nicely factorized expression.
It begins from $\lambda$, because the contribution of $N=0$ (Alexander polynomial)
is nullified by the factor $D_R$.

This answer can be easily generalized to other torus knots and other single-trace averages,
for instance, for 2-strand torus knots one has
\be
   {\cal Z}^{{\rm Torus}_{2,n}}_{1}
= \frac{\lambda q^n (q^n-\lambda)}{  (1-q^{n-2}\lambda)(1-q^{n }\lambda)(1-q^{n+2}\lambda)}
=
% ???
\lambda q^{2 n} \cdot
\frac{(q^{- n} \lambda; q^2)_\infty}{(q^{- n + 2} \lambda; q^2)_\infty}
\cdot
\frac{(q^{n + 4} \lambda; q^2)_\infty}{(q^{n - 2} \lambda; q^2)_\infty}
\ee
with the standard notation for the $q$-Pochhammer symbol
\begin{align} \label{eq:q-pochhammer}
  (a; q)_\infty = \prod_{k=0}^\infty (1 - q^k a)
\end{align}

\section{Failures beyond torus fundamental HOMFLY}

In this section we list the cases, when factorization does not occur already in the
simplest HZF -- for the knot polynomial in the fundamental representation.
What unifies these case -- for torus and non-torus knots -- is the lack of
{\it explicit} eigenvalue matrix model, which converts characters into knot polynomials.

\subsection{Reduced polynomials and Alexander polynomials}

If we define averages with the help of reduced polynomials,
things would be different.
In particular at $N=0$ we get at the r.h.s. the Alexander polynomials,
which depend on representation only through the size of the
single-hook diagram \cite{
  paper:MM-eigenvalue-conjecture-and-colored-alexander-polynomials
}
\be
\left<P_k\right>^{\cal K}_{N=0}
= \sum_{i=0}^k (-)^i {\cal A}^{\cal K}_{[k-i,1^{i}]}(q)
=  {\cal A}_{[1]}(q^k) \cdot \left(\sum_{i=0}^k (-)^i\right)
= {\cal A}_{[1]}^{\cal K}(q^k) \cdot \delta_{k,{\rm odd}}
\ee
From \cite{
  %DMMSS
  paper:DBMMSS-superpolynomials-for-torus-knots-from-evolution-induced-by-cut-and-join-operators
} for torus knots
\be
{\cal A}^{{\rm Torus}_{m,n}}_{[1]}
= \frac{(q-q^{-1})(q^{mn}-q^{-mn})}{(q^{m}-q^{-m})(q^{n}-q^{-n})}
\ee
so $\left<P_k\right>^{\cal K}_{N=0}$ does have the structure \eqref{statement},
but this does \text{not} generalize to $N \neq 0$.

\noindent Namely, the needed average would be
\be
\overline{\left< {\rm Schur}_R\right>^{{\rm Torus}_{m,n}}_{SL(N)}} =
\sum_{Q\in m|R|} q^{\frac{2n\varkappa_Q}{m}}\cdot C_R^Q \cdot
\frac{{\rm Schur}_Q\left\{\frac{[Nk]}{[k]}\right\}}{{\rm Schur}_R\left\{\frac{[Nk]}{[k]}\right\}}
\label{aveSchurred}
\ee
and the corresponding Harer-Zagier formula does not factorize already for trefoil
\be
\overline{{\cal Z}^{{\rm Torus}_{2,3}}_{1}}
= -\frac{q^2+q^{-2}}{1-q^4\lambda} - \frac{1}{1-q^2\lambda}
= \frac{q^2-1+q^{-2}-\lambda q^2}{(1-q^4\lambda)(1-q^2\lambda)}
\ee
Factorization, however, occurs for the $q$-derivative
\be
{\cal Z}^{{\rm Torus}_{2,3}}_{1} = \frac{\overline{{\cal Z}^{{\rm Torus}_{2,3}}_{1}}(\lambda q) -
\overline{{\cal Z}^{{\rm Torus}_{2,3}}_{1}}(\lambda/q)}{q-q^{-1}}
\ee
but there is no clear reason for it, besides pure technical on one hand
and explicit existence of the TBEM  eigenvalue model on the other hand.

\subsection{First non-torus knot: figure 8}

It is easy to check that at least some other knots in the fundamental representation do not possess
such HZ factorization property.
For example, for the figure-eight knot $4_1$ the Harer-Zagier functions
(for both normalized and non-normalized HOMFLY-PT) do not factorize:
\be
H^{4_1}_{[1]} = & \left(A^2+\frac{1}{A^2}-q^2-\frac{1}{q^2}+1\right)
\\
\overline{{\cal Z}^{4_1}_{1}}
= & -\frac{\lambda^2 q^4+\lambda q^8-2 \lambda q^6-2 \lambda
  q^2+\lambda-q^6+3 q^4-q^2}{(\lambda-1) q^2 \left(\lambda-q^2\right)
  \left(\lambda q^2-1\right)}
\nn \\
{\cal Z}^{4_1}_{1} = \frac{\overline{{\cal Z}^{4_1}_{1}}(\lambda q) -
\overline{{\cal Z}^{4_1}_{1}}(\lambda/q)}{q-q^{-1}}
= & \frac{\lambda \left(\lambda^2 q^5+\lambda q^{10}-\lambda
  q^8-\lambda q^6-\lambda q^4-\lambda q^2+\lambda+q^5\right)}{q
  (\lambda-q) (\lambda q-1) \left(\lambda-q^3\right) \left(\lambda
  q^3-1\right)}
\ee

\subsection{Superpolynomials}

Likewise, the torus superpolynomials are also not suitable:
\be
P_{[1]}^{(2,3)} \sim A^2(A^2 t^2 + A^2 q^{-2} - 1) \frac{A - A^{-1}}{t-t^{-1}} \ \ \ \ \ \stackrel{A=t^N}{\Longrightarrow} \
\ee
\vspace{-0.5cm}
{\fontsize{9pt}{0pt}{\be
Z_1^{(2,3)} = \frac{1}{t-t^{-1}} \left( \frac{t^2 +q^{-2}}{1-\lambda t^5} -   \frac{t^2+1+q^{-2}}{1-\lambda t^3} + \frac{1}{1-\lambda t} \right)
= \frac{t^2}{q^2} \frac{\lambda\left(t^4q^2+t^2-q^2-\lambda t^3\right)}{(1-\lambda t)(1-\lambda t^3)(1-\lambda t^5)}
\ \stackrel{t=q}{\longrightarrow} \
\frac{\lambda q^3 (q^3-\lambda)}{(1-q\lambda)(1-q^3\lambda)(1-q^5\lambda)}
\nn
\ee}}
Perhaps, for things to work, one needs to modify ($t$-deform) the prescription \eqref{eq:laplace-transform}
on how to do the Laplace transform. However, at this point this is pure speculation,
and this line of development will be pursued elsewhere.

\subsection{Kauffman polynomials}

Kauffman polynomials are the analogues of HOMFLY-PT for the $SO(N)$ gauge groups.
They are also desribed by an analog of Rosso-Jones formula
\begin{align}
  K_R^{\text{Torus}_{n,m}} = q^{m n (N-1) \varkappa_R} q^{m n |\lambda|}
  \sum_{|\nu| \leq n |R|} b^{\nu}_{R, n} q^{-\frac{m}{n} (N-1) \varkappa_\nu} q^{-\frac{m}{n} |\nu|} d_\nu
\label{Kauftor}
\end{align}
where  $d_\nu$ is the quantum dimension of $SO(N)$ representation, corresponding to diagram $\nu$
and coefficients $b^{\nu}_{R, n}$ are the direct analog of Adams coefficients. The details
of their calculation can be found in
\cite{paper:stevan}.

In topological framing unreduced fundamental Kauffman satisfies
\be
K_{\Box,{\rm red}}^{\cal K}-1\ \vdots\  (A-q)(Aq+1)(A+1)(A-1)
\ee
From (\ref{Kauftor}) one can deduce Kauffman polynomial for trefoil in fundamental representation
\begin{align}
K_{\Box,{\rm red}}^{\text{Torus}_{2,3}}
= -A^2\cdot \frac{A^3(q^3-q)+A^2(q^4 - q^2+1) - A(q^3-q)-(q^4 +1)}{q^2}
\label{Kaufund}
\end{align}
but its Laplace transform {\it does not} factorize.
Moreover, this time it is not cured by multiplication with any reasonable functions of $A$,
even different from dimension.
One can wonder what this means from the point of view of our
{\it matrix model -- factorization} hypothesis.
The answer is that the $SO(N)$ models are assocaited with $\beta=2$ rather than $\beta=1$
(rouhgly speaking, $2\beta$ is the power of Vandermonde-like factor in the measure).
TBEM formula with $\beta=1$ is easily generalized to simply-laced groups,
which includes $D$, but not $B$ --
while (\ref{Kaufund}) for Kauffman invariant unifies even and odd $N$ through $A=q^{N-1}$ --
and the unifying matrix model shooul gave $\beta=2$.
This puts Kauffmann example into intermediate position, and once again calls for
the proper understanding in group theoretic terms
of the Laplace transform, and its proper generalization.

\bigskip

To summarize, we illustrated the distinguished role of HOMFLY-PT polynomials in torus family
from the point of view of factorizability of HZF:
any deviation seems to violate it.
Instead, as we claim in the next section, for {\it torus} HOMLY-PT factorization is  true
for all single-trace HZF $\left<p_k\right>$, not only for   $\left<p_1\right>$.

\section{Other single-trace correlators}

For higher representations of $SU(N)$ the issue of framing becomes important,
because we get a linear combination of different representation in $R^{\otimes m}$,
which transform differently under a change of framing.
%As it turns out, the
Factorization takes place in the so-called \textit{spectral} (or {\it vertical}) framing,
%This framing
which was used by Rosso and Jones in their original formula for colored HOMFLY
\cite{
  paper:RJ-torus-knots-paper
}
and is exactly the framing, assumed in the Tierz-Brini-Eynard-Marino matrix model
\cite{paper:T-soft-matrix-models-and-chern-simons-partition-functions,
  paper:BEM-torus-knots-and-mirror-symmetry
},
and in which the Ooguri-Vafa partition function is nicely expressed in terms of the free-fermion formalism
\cite{
  paper:DBPSS-combinatorial-structure-of-colored-homfly-pt-polynomials-for-torus-knots,
  paper:DBKPSS-topological-recursion-for-the-extended-ooguri-vafa-partition-function-of-colored-homfly-pt-polynomials-of-torus-knots
}.

Namely, the Rosso-Jones formula in this framing reads
\begin{align}
  {\cal H}_R^{{\rm Torus}_{m,n}} (A, q)
  = A^{n |R|} \sum_{Q\in m|R|} q^{\frac{2n\varkappa_Q}{m}}\cdot C_R^Q\cdot
  {\rm Schur}_Q \left \{p_k = \frac{A^k - A^{-k}}{q^k - q^{-k}} \right \},
 \label{torfor1}
\end{align}
where $C_R^Q$ are, again, Adams coefficients from (\ref{adams}).
Note that vertical framing explicitly breaks the $n \leftrightarrow m$ symmetry.

\bigskip

With this convention, the Laplace transform of single-trace correlator becomes nicely factorized:
\begin{align}
\!\!\!\!
  \boxed{
    \mathcal{Z}_d^{\text{Torus}_{n,m}}\!
    = \lambda q^{d^2 nm}\cdot
    \frac{\prod_{i = 0}^{d m - 2}  \Big(1 - q^{d (- n - m) + 2 + 2 i} \lambda \Big)}
    {\prod_{i = 0}^{d m}  \Big(1 - q^{d(n-m) + 2 i}\Big)}
    = \lambda q^{ d^2 n m }\cdot \frac{(\lambda q^{d (- n - m) + 2}; q^2)_\infty}{(\lambda q^{d (- n + m)}; q^2)_\infty}
    \frac{(\lambda q^{d (n + m) + 2}; q^2)_\infty}{(\lambda q^{d (n - m)}; q^2)_\infty}
  }
\nn
\end{align}
where we once again use the $q$-Pochhammer symbols \eqref{eq:q-pochhammer}.

\bigskip

This factorization formula for the Laplace transform in $N$ of the peculiar combination of torus HOMPLY-PT
polynomials is the main result of this paper.
%One may transform it to a more or less nice form with slight modification
%of the framing, yet, in the absence of \textit{other} examples that also work,
%there seem to be no reason to prefer one such framing over another.

The proof of this identity is calculational and combinatorial, very similar in spirit to that of
\cite{
  % MPSh
  paper:MPSh-quantization-of-harer-zagier-formulas
}.
The main point is seen already at $q=1$: according to (\ref{adams}), in this case we deal just with the dimension
of representation ${\rm Schur}_R\{N\}$, whose Laplace transform is factorizable only for single-hook $R$,
while factorization is lost beyond one-hook (and thus for multi-trace averages), e.g.
$\sum_N \lambda^N\cdot {\rm Schur}_{[3,3]}\{N\}
= \frac{\lambda^2(\lambda^2+3\lambda+1)}{(1-\lambda)^7}$.
Moreover, the single-hook factorization
%persists after a $q$-deformation, but is violated by the
is preserved by the
$q$-deformed Adams rule (\ref{adams}) and
the action of $\hat W$ operator
\cite{%MSh_w
  paper:MSh-generation-of-matrix-models-by-w-operators
}, which is responsible for the factor $q^{-\frac{2n\varkappa_Q}{m}}$
in (\ref{torfor1}).
%The action of $\hat W$ operator \cite{MSh_w}, which is responsible for the factor $q^{-\frac{2n\varkappa_Q}{m}}$
%in (\ref{torfor1}),
%corrects this violation
%%%%corrects the violation of this property by the $q$-deformed Adams rule
%%%%preserves this factorization property
%%%%under $q$-deformation --  this
This is technically straightforward:
\begin{align*}
{\cal Z}^{\rm{Torus}_{n,m}}_d \
& = \sum\limits_{N = 0}^{\infty} \ \lambda^N \ \sum\limits_{i = 0}^{d-1} \ (-1)^i \
{\cal H}^{\rm{Torus}_{n,m}}_{[d-i,1^i]}(q,q^N)
\\
& = \sum\limits_{N = 0}^{\infty} \ \lambda^N \  q^{Ndn} \sum\limits_{Q \vdash dm} \ q^{2\frac{n}{m}\varkappa_Q} \
\underbrace{\left( \sum\limits_{i = 0}^{d-1} \ (-1)^i \ C^{Q}_{[d-i,1^i]} \right)}_{\sum\limits_{L=0}^{d m - 1} \ (-1)^L \ \delta_{Q, [dm-L,1^L]}} \ {\rm Schur}_{Q}\left(p_k = \dfrac{q^{Nk}-q^{-Nk}}{q^k-q^{-k}} \right)
\\
& = \sum\limits_{N = 0}^{\infty} \ \lambda^N \ q^{Ndn} \sum\limits_{L = 0}^{dm-1} \ (-1)^L \ q^{2\frac{n}{m}\varkappa_{[dm-L,1^L]}} \ {\rm Schur}_{[dm-L,1^L]}\left(p_k = \dfrac{q^{Nk}-q^{-Nk}}{q^k-q^{-k}} \right)
\\
& =
\sum\limits_{L = 0}^{dm-1} \ (-1)^L \ q^{2\frac{n}{m} \frac{dm(dm - 2L - 1)}{2} } \dfrac{1}{q^{dm}-q^{-dm}} \ \dfrac{\sum\limits_{N = 0}^{\infty} \ \lambda^N \ q^{Ndn} \ \prod\limits_{s=0}^{dm-1} (q^{N-L+s}-q^{-N+L-s})}
{\prod\limits_{s=1}^{dm-L-1} (q^{dm-L-s}-q^{-dm+L+s})
\prod\limits_{s=0}^{L-1} (q^{L-s}-q^{-L+s})
}
\\
& = \dfrac{\prod\limits_{s=1}^{dm-1}(q^s-q^{-s})}{\prod\limits_{s=0}^{dm}(1-q^{d(n-m)+2s}\lambda)} \sum\limits_{L = 0}^{dm-1} \ \dfrac{(-1)^L \ (\lambda q^{dn})^{L+1} \ q^{dn(dm - 2L - 1)}}{\prod\limits_{s=1}^{dm-L-1} (q^{dm-L-s}-q^{-dm+L+s})
\prod\limits_{s=0}^{L-1} (q^{L-s}-q^{-L+s})
}
\\
& = \lambda \ q^{d^2 nm} \ \dfrac{\prod\limits_{s=0}^{dm-2}(1-q^{d(-n-m)+2+2s}\lambda)}{\prod\limits_{s=0}^{dm}(1-q^{d(n-m)+2s}\lambda)}
\end{align*}
The main point here is the double application of the projection rule (\ref{projk}) 
to the definition (\ref{adams}) of Adams coefficients:
\be
\sum_{R} (-)^i\cdot\delta_{R,[d-i,1^i]}\cdot 
\left(
\sum_{Q\vdash m|R|} C^Q_R \cdot{\rm Schur}_R\{p_k\} \ \stackrel{(\ref{adams})}{=} \ {\rm Schur}_R\{p_{km}\}
\right)
\ \ \Longrightarrow \nn 
\ee
\vspace{-0.4cm}
\be 
\!\!
\sum_Q \left(\sum_{i} (-)^i \,C^Q_{[d-i,1^i]}\right) {\rm Schur}_Q\{p_k\} 
= \sum_i (-)^i {\rm Schur}_{[d-i,1^i]}\{p_{km}\} \ \stackrel{(\ref{projk})}{=} \  p_{dm}
 \stackrel{(\ref{projk})}{=} \ 
\sum_L (-)^L\cdot{\rm Schur}_{[dm-L,1^L]}\{p_k\}
\ \ \Longrightarrow \nn 
\ee
\vspace{-0.4cm}
\be 
\sum_{i=0}^{d-1} (-)^i \, C^Q_{[d-i,1^i]} = \sum_{L=0}^{dm-1} (-)^L\,\delta_{q,[dm-L,1^L]}
\ee
Despite apparent simplicity of this calculation,
a conceptual proof is highly desirable, 
applicable to the whole variety of $\beta=1$ matrix models.
%of this fact
%is highly desirable.

\section{Conclusion}

In this paper we proposed to view the factorization of the Laplace transform
of single-trace average as an alternative manifestation of \textit{superintegrability}.

Remarkably, this factorization turns out to be present for torus knots' HOMFLY-PT polynomials,
where the eigenvalue model (and free-fermion representation) is explictly known,
but the very naive attempts to observe similar factorization
for slight deformations of this setting: to superpolynomials, to other knots and even to Kauffman polynomials
-- all fail.

This seems to suggest, that the problem of finding proper matrix models for families of knot polynomials
is more tricky and rigid than was first thought.
One of the ways around this situation would be to relax the prescription
\begin{align}
  \left\langle character \right\rangle = \text{knot polynomial}
\end{align}
in some yet unknown way.
As was recently demonstrated \cite{
  paper:MM-superintegrability-of-kontsevich-matrix-model
} similar broadening of the point of view
can be very fruitful in discovering new character expansion formulas.

Another possibility would be to understand the group-theoretic meaning of the Laplace transform,
and to adjust it, accordingly. At the moment this part of the Harer-Zagier construction seems completely \textit{ad hoc}.

There is a number of possible directions/questions to pursue, which we would like to point out
\begin{itemize}
\item What is the formula for the double-trace correlators? Is it, in some sense, similar
  to the one for $q$-deformed Hermitian Gaussian matrix model?
\item HOMFLY-PT polynomials for torus knots have a well-known generalization from $S^3$ to Seifert spaces.
  The knot matrix model is known for this case. Does the Harer-Zagier factorization persist as well?
\item Superpolynomials, which we considered in this paper, are not Khovanov-Rozansky polynomials.
  Instead, they coincide for large enough $N$ (which is knot-dependent).
  Therefore, the Laplace transformed sums for superpolynomials and actual KR polynomials differ
  by some polynomial in $\lambda$. Can it be, that this polynomial transforms non-factorizable
  Harer-Zagier function into factorizable?
\item Last but not least, there is a question about the relation of superintegrability (in the form of Harer-Zagier factorization)
  to other well-known, and undergoing rapid development, knot-theoretical structures: the knots-quivers correspondence
  \cite{paper:K-quivers-for-3-manifolds-the-correspondence,
    paper:EGGKSS-z-at-large-n-from-curve-counts-to-quantum-modularity,
    paper:EKL-multi-cover-skeins-quivers-and-3d-n-dualities,
    paper:KRSS-knots-quivers-correspondence
  }
  and theory of q-Virasoro localization
  \cite{paper:CLZ-on-matrix-models-and-their-q-deformations,
    paper:NNZ-q-virasoro-modular-double-and-3d-partition-functions,
    paper:NZ-q-virasoro-constraints-in-matrix-models}.
\end{itemize}

We hope to address some, or all of these questions in future.

\section*{Acknowledgements}

The work is partly supported by RFBR grants  19-02-00815 (A.M., Sh.Sh.),
19-51-18006\_Bolg-a (A.M.),  19-01-00680 (A.P.)
and by the joint {RFBR-MOST} grant  21-52-52004 (A.M., A.P.)

\bibliographystyle{mpg}
\bibliography{references_hz-torus-knots}

\end{document}